\input mtexsis
\tenpoint
\def\Tbf{\fourteenpoint\bf}
\headline={\hfil}
\footline={\hfil\folio\hfil}
\title
{\Tbf{Potentiality, Entanglement and Passion-at-a-Distance$^{\dag}$}}
\endtitle
\vskip 0.03in
\center
Joy ~Christian
\smallskip\tenpoint
{\it Wolfson College, University of Oxford, Oxford OX2 6UD, United Kingdom}
\tenpoint
{\it Electronic address} : {\it joy.christian@wolfson.oxford.ac.uk}
\endcenter

\vskip 0.14in
\baselineskip 0.51cm

\parindent 0.00cm

This essay is a review of
{\it Potentiality, Entanglement and Passion-at-a-Distance: Quantum Mechanical
Studies for Abner Shimony, Volume Two}; Boston Studies in the
Philosophy of Science, Volume 194; edited by Robert S. Cohen, Michael
Horne and John Stachel (Dordrecht: Kluwer Academic Publishers,
1997), xi + 268 pp., ISBN 0-7923-4453-7, Cloth
${\hbox{\it\char36}}$81.00.
{\parindent 0.00cm\parskip -0.50cm
\baselineskip 0.53cm\footnote{}{\tenpoint{\hang ${\dag}$
To appear in
{\sl Studies in History and Philosophy of Modern Physics} (1999).\par}}}

\vskip 0.14in
\parindent 0.55cm

Among our two most basic physical theories quantum mechanics enjoys a special
status, thanks to its distinctive and counterintuitive characteristics such as
`potentiality', `entanglement' and `non-locality'. To be sure, general
relativity is its worthy rival in beauty, internal coherence and unprecedented
empirical precision, but in its ability to arouse acrimonious debates over
a possible {\it interpretation} of its formalism quantum theory stands second
to none. There is, of course, no disagreement over the set of rules embedded in
the quantum formalism for predicting experimental outcomes. But these rules
are disconcertingly silent on how one should picture the underlying quantum
{\it reality}. And when an attempt is made to provide such a picture, one is
faced with some curious and fascinating metaphysical dilemmas never before
encountered in the history of physics. The nineteen essays
in {\it Potentiality, Entanglement and Passion-at-a-Distance} -- a Festschrift
dedicated to Abner Shimony, who is renowned for his pioneering contributions to
the subject -- cover a wide range of issues related to these dilemmas. The
contributions -- almost all of which are by eminent workers in
the field -- are not intended for the uninitiated, and require sophistication
in physics well beyond the undergraduate level.

\parskip 0.17cm

The neo-Aristotelian notion of quantum-mechanical {\it potentiality} as a novel
metaphysical modality of nature -- situated between mere logical possibility
and {\it bona fide} actuality -- was
favoured by Heisenberg, and has been exuberantly endorsed by
Shimony (1978, 1998).
According to this notion, the quantum-mechanical statevector represents a
network of potentialities governed by the well-known {\it linear}
Schr\"odinger dynamics. A potentiality for a physical property is far more
than a mere (objective) indefiniteness of that property, since it also includes
further structure such as non-classical correlations between quantum
constituents (Shimony 1998).
By contrast, {\it actuality} --  `emerging' ontologically
via a controlled or uncontrolled
`act of measurement' -- embodies definiteness, resulting from a loss of quantum
correlations. In terms of these notions, the quantum-mechanical measurement
problem is simply an acute problem of unequivocally understanding the
phenomenologically prolific {\it transitions} from potentialities to
actualities -- i.e., understanding the ubiquitous transitions between these
two completely distinct metaphysical modalities.

If potentialities are as real as actualities, then can one observe a network of
such potentialities? In more familiar terms:
Is the quantum state observable? This is
the question asked in two of the contributions to the Festschrift. Aharonov and
Vaidman pose this question very differently from Busch. They introduce a
variant of the notion of `protective measurement' (previously introduced by
them in collaboration with Anandan) to `observe' the quantum state of a single
system, and generalize it to Aharonov's two-vector formulation of quantum
mechanics. Busch, on the other hand, explores the issue within the general
representation of observables as positive operator valued measures, and
concludes that ``the quantum state is not an observable [in the usual
quantum-mechanical sense], but [nevertheless] not unobservable''.

One of the most profound manifestations of potentiality is, of course,
quantum entanglement, which entails that the quantum-mechanical `whole' is
profusely and quantitatively more than simply `the sum of its parts'.
Since not all quantum-mechanical states are entangled states (some being
product states), an immediate question that arises about a given quantum state
is whether or not it is entangled. More precisely: ``Is it algorithmically
decidable whether a given quantum-mechanical state is entangled or not?'' And
``if not, can we effectively decide whether or not a state is within [a given
Hilbert-space distance] ${\varepsilon}$ of a product state?'' Amusingly enough,
as Myrvold reports in his contribution to the Festschrift, the first of these
two ``decision problem[s] for entanglement'' was posed to Shimony by Aspect in
a dream! The second problem was subsequently posed by Shimony (fully
conscious!) to Myrvold, who in his contribution reaches the conclusion that
the question ``Is there a product state within [a given Hilbert-space distance]
${\varepsilon}$ of ${\psi}$?'' is almost decidable, but not quite. That is to
say, for a given Hilbert-space vector ${\psi}$ and a real number
${\varepsilon}$, there is an algorithmic procedure that gives an affirmative
answer to this question if there happens to be a product state (say ${\chi}$)
less than a distance ${\varepsilon}$ away from ${\psi}$
(i.e., if ${||\psi-\chi||<\varepsilon}$), and a negative answer if all product
states are further than the distance ${\varepsilon}$ from ${\psi}$. However,
not surprisingly (not, that is, to an
expert in orthodox computability theory),
the algorithm fails to terminate if the distance of the nearest product state
from ${\psi}$ is {\it exactly}
equal to ${\varepsilon}$. In other words, the question
whether ${\psi}$ is precisely a product state (i.e., whether
${||\psi-\chi||=\varepsilon = 0}$) is effectively undecidable!

In addition to this analysis by Myrvold, about a third of the contributions in
the Festschrift are devoted to quantum entanglement and its implication of
`non-locality'. Since the non-locality harboured within quantum entanglement
seems devoid of any `action-at-a-distance' (it does not lead to sending a
`signal' faster than light), Shimony has humourously used the expression
`passion-at-a-distance' to describe it: hence the appearance of this locution
in the title of the Festschrift. At times, borrowing a political aphorism,
Shimony has also referred to this curious state of affairs as `peaceful
coexistence' between quantum non-locality and the causal requirements of
special relativity. There are some remarkable contributions in the Festschrift
which examine this issue of `peaceful coexistence'. Particularly noteworthy
is the analysis by Popescu and Rohrlich, who, following Shimony, raise one of
the most daring questions of all: ``Why is quantum mechanics what it is?''
In particular: Is it possible for non-locality and causality to peacefully
coexist in any other theory besides quantum mechanics? Curiously enough, by
constructing some explicit models, Popescu and Rohrlich demonstrate that
quantum mechanics is {\it not} the only theory that is hospitable to both
non-locality and causality; although quantum-mechanical non-locality is
certainly different in character from any non-locality that could be
accommodated in the general framework of classical physics.

Of course, the historical roots of quantum non-locality can be
traced back (at least) to the pioneering argument put forward by Einstein,
Podolsky, and Rosen (EPR) in 1935. It is then somewhat surprising,
considering the purported threat of action-at-a-distance central to this
argument, that so little attention is given to its special-relativistic
reformulation. One reason for this neglect is perhaps the
difficulties involved in translating the argument into a relativistic context.
In their contribution to the Festschrift, Redhead and La Rivi\`ere discuss the
problematics of such a translation, paying particular attention to the need for
a relativistic modification of the {\it reality criterion} of EPR. They note
that precisely in this reality criterion a blatantly nonrelativistic assumption
of absolute time ordering among the measurement events occurring in the two
branches of an entangled system enters the EPR argument, necessitating a
modified criterion for the relativistic systems. One such relativistic reality
criterion for the attribution of elements of reality that is {\it not}
contingent on the time ordering of the actualizations of potentialities has
been previously proposed by Ghirardi and Grassi. However, Redhead and
La Rivi\`ere argue that the Ghirardi-Grassi criterion is flawed by an
ambiguously stated locality principle and an assumption of determinism hidden
in the use of counterfactuals, undermining their claimed peaceful
coexistence between quantum mechanics and special relativity. Elsewhere in the
Festschrift Ghirardi provides a vigorous defence for the Ghirardi-Grassi
criterion against these charges by Redhead and La Rivi\`ere. Judging from this
stimulating exchange it appears, however, that a
satisfactory relativistic formulation of the EPR argument is still very much
an open question.

Returning to the notion of quantum-mechanical potentiality, it is worth noting
that the notion not only lends metaphysical
clarity to quantum mechanics, but also allows a convenient classification of
different (realist) approaches to the interpretational problems infesting the
theory (Shimony 1998). These approaches fall roughly into three broad
categories, which -- divulging my own order of preference on the matter --
I propose to call: (1) regressive approaches, (2) fallacious approaches, and
(3) progressive approaches.

The first among these downplay the metaphysical
innovations brought about by quantum mechanics; in particular, they downplay
the concepts of indefiniteness and potentiality by attempting to reduce them --
by hook or by crook -- to mere reflections of ignorance. A prime
example of such a regressive approach which tries to undermine the conceptual
revolutions brought about by quantum theory is the `hidden variables' programme
(including the de Broglie-Bohm theory). It remains to be seen whether any such
counterrevolutionary
approach is eventually successful in reducing quantum mechanics
(especially relativistic quantum field theory) to a mere glorified version of
(classical) statistical mechanics.

The second category of approaches
-- to their credit -- embrace potentialities wholeheartedly, but deny
actualizations of these potentialities. They acknowledge {\it appearances} of
such actualizations, but view these appearances as purely phenomenological
aspects of the world. The examples here are `the many-worlds interpretation'
of quantum mechanics and various decoherence theories. Such an approach
takes the formalism of quantum theory too seriously, prematurely extrapolating
it to {\it all} physical scales
(from microscopic to cosmological), and thereby commits -- as Whitehead would
have put it -- `the fallacy of misplaced concreteness'. Indeed, in the light of
the extraordinary specialness of Planck-scale physics{\parindent 0.37cm
\parskip -0.50cm \baselineskip 0.53cm\footnote{*}{\ninepoint{\hang
The Planck scale is indeed quite dramatically
more special compared to any other
better-understood scale in physics, since, near it (e.g., near the big bang or
a black hole), some {\it evenhanded} interplay between the fundamental tenets
of quantum mechanics and general relativity is undoubtedly taking place. What
is more, understanding this interplay would almost certainly necessitate
some radically unorthodox ideas, since the non-dynamical spacetime structure
axiomatic to quantum theories is utterly anathematic to the very essence of
general relativity,
with its {\it dynamical} picture of spacetime. In the light of these facts, and
because the conventional distinction between `microscopic' and `macroscopic'
ceases to be meaningful near the Planck scale, it is rather presumptuous -- if
not naive -- to maintain that only general relativity breaks down near
this unique physical scale while quantum mechanics escapes it unscathed.
For further discussion on these and related issues see (Christian
1998).\par}}},
the {\it a priori} extrapolation by some {\it seventeen} orders of magnitude
(e.g., in the mass scale)
required by these approaches to maintain the universality of the quantum
formalism can hardly be viewed as rational.

Finally, the third category of approaches, which actually
go beyond the confining formalism of quantum mechanics (e.g., a Pearle-GRW-type
theory), not only recognizes the metaphysical innovations brought about by
quantum mechanics, such as potentialities, but also takes the proliferation of
actualities in the world to be more than a mere phenomenological experience.
That is, these approaches put actuality on a par with potentiality, and take
both to be equally genuine ontological attributes of the world. Unfortunately,
although they make some headway towards explaining the process of
the actualization of potentialities (but see below), as yet there is no
empirical support for these innovative approaches. Furthermore, they face
immense theoretical and conceptual difficulties. Nevertheless, since
a reconciliation between quantum mechanics and general relativity -- the other
(often unfairly neglected) pillar of ${20^{th}}$ century physics -- is as yet
beyond the horizon, and such a reconciliation will almost certainly
bring about a major revolution in notions as basic as causality and time,
the proponents of this third category are more than justified in looking beyond
the straitjacket of orthodox quantum theory. Indeed, the problem of the
actualization of potentialities is best viewed not merely as a problem to be
solved within the framework of quantum mechanics, but rather as a glorious
opportunity to go beyond the confines of quantum theory
in order to comply with the ineluctable demands of general relativity.

In this Volume Two of the Festschrift, there are no articles
adhering to a regressive approach belonging to category (1) above; there
are a few in Volume One, however. The category (2) of fallacious approaches
fares better in this
respect, in that there are several articles in this category; although
there is no explicit adherent of `many-worlds' in these articles. Refreshingly
enough, however,
there is at least one essay -- that by Ghirardi and Weber -- which explicitly
advocates going beyond the bounds of quantum mechanics, {\it \`a la} the
Pearle-GRW programme, and hence belongs to the category (3) of progressive
approaches. Going beyond quantum mechanics is, of course, easier said than
done, and the task Ghirardi and Weber set out to accomplish is far from
trivial.
In the Pearle-GRW programme we are asked to accept an {\it ad hoc} assumption:
``besides the standard  [quantum-mechanical] evolution, physical systems are
subjected to spontaneous localizations occurring at random times and affecting
their elementary constituents.'' If this assumption is accepted without
reservations, then we are promised the much wanted `holy grail': a universal
dynamics which remains valid for both micro and macro systems, is conducive to
an `individual interpretation' for these systems, and -- most importantly --
provides an ``account for the emergence, at the appropriate level, of definite
properties of individual physical systems''. Remarkably enough, as is now
well-known, the Pearle-GRW programme does succeed in achieving the first
part of this goal. Unfortunately, however, due to the infamous `tail problem',
the attempt miserably fails when it comes to providing a convincing account for
the {\it ontic} emergence of definite properties.

Ghirardi and Weber mount a desperate but -- at least in my view --
ultimately unsuccessful attempt (as
does Pearle differently in Volume One) to convince us that the `tail problem'
is not really a problem at all, if the Pearle-GRW model is interpreted
`correctly'. The issue is really quite simple: their unified dynamics reduces
all superposed states of a sufficiently macroscopic body,
such as ${\alpha\,\ket{\rm here}} + {\beta\,\ket{\rm there}}$,
to {\it almost} disjoint states -- either
${\ket{\rm here}}$ or ${\ket{\rm there}}$, but not quite, because the unwanted
complex coefficient of such a superposition, say ${\beta}$, is -- albeit
extremely small (${\beta < < < \alpha}$) -- never exactly zero, leaving a
small but infinite `tail' in the wrong part of the wavefunction. Although this
is good enough `for all practical purposes' (which is what the Ghirardi-Weber
attempt of reinterpretation boils down to), from the metaphysical perspective
the `tail' is disastrous. For, clearly, as long as the `tail' remains in either
explicit or implicit form, a Pearle-GRW-type approach makes not an {\it iota}
of improvement over the standard quantum measurement theory with respect to the
metaphysical problem of the actualization of potentialities. Since no amount
of reinterpretation of Pearle-GRW formalism can make the unwanted coefficients
vanish, the efforts of Pearle in Volume One and that of Ghirardi and Weber in
the present volume are distressingly futile, leaving Schr\"odinger's Cat
to remain as haunting as ever!

The `tails' in the Pearle-GRW programme are but a symptom of a much deeper
problem. Let us not forget
that these models begin with an {\it ad hoc} assumption of
`spontaneous localizations occurring at random times'. This assumption is
purely phenomenologically motivated, and the authors offer little physical
underpinning -- if any -- for this awkward process. To be sure, one comes
across some quasi-physical notions in the literature such as `quantum
fluctuations of spacetime' offered as a physical reason behind the assumption,
but in the absence of a consistent quantum theory of gravity such
ill-understood (if not entirely meaningless) notions do little to comfort
one's unease with the crudeness of the process. Clearly, a more honest approach
would be to view the `tail problem' as symptomatic of a much deeper physical
problem -- that of understanding the {\it actual} physics
underlying the mechanism of quantum state reduction. In this context, the
Pearle-GRW approach should be contrasted with the (slightly) related proposal
by Penrose (1996, Christian 1998), who at least identifies the
fundamental {\it physical} culprit, the {\it raison d'\^etre} for state
reduction -- namely the irreconcilable conflict between principles of
superposition and general covariance, albeit as yet without a detailed theory. 

That gravity should be responsible for quantum state reduction is an
intriguing proposition, gaining strength from the fact that {\it all} attempts
to reconcile gravity with the quantum formalism have failed. There are two
contributions in the Festschrift which touch upon some of the acute issues
faced by any attempt to `unify' general relativity and quantum mechanics. In a
thought provoking essay, Anandan highlights a parallel between Einstein's
efforts to create a {\it relativistic} theory of gravity and the task we face
of creating a `{\it quantum}
theory of gravity'. Drawing lessons from this observation, he proposes quantum
versions of the principles of equivalence and general covariance. Howard, on
the other hand, after distinguishing between the two possible senses of
non-locality discussed above, notices that, unlike the `peaceful coexistence'
({\it \`a la} Shimony)
between the quantum-mechanical non-locality and the causal requirements
of special relativity, there is no peaceful coexistence between the
event-ontologies of {\it general} relativity and quantum mechanics, and
sketches a programme for `quantum gravity'
which could put the two theories in ontological
harmony. However, he seems to be completely oblivious of Penrose's twistor
programme for `quantum gravity' (now well in its thirties), which is not only
conceptually and mathematically far more sophisticated than Howard's embryonic
suggestion, but also -- since `non-locality' is an intrinsic feature of the
twistor space --
goes a long way towards making `peace' between the two ontologies.

Out of the nineteen contributions in the Festschrift, I have discussed but a
handful, and even those only impressionistically.
There is a lot to choose from. In particular, there
are contributions from some of the most eminent workers in the field such as
d'Espagnat, Hardy, Mermin, Mittelstaedt, Peres, Primas, Rimini,
Stachel, and Stein, and it is impossible to do justice
to their varied viewpoints in this limited space. Although some of the
articles in the Festschrift are quite demanding -- and at times rather
trite -- I highly recommend the collection as providing much food
for thoughts on quantum dilemmas.

Finally, at the end of the Festschrift we are given a bibliography of
Shimony's publications starting from 1947. The length of this bibliography is
impressive enough, but what is even more impressive is the uncanny versatility
of Shimony's scholarly mind. For instance, in addition to his enviable
contributions in pure philosophy, philosophy of physics, theoretical physics,
and experimental physics, we come across his work on philosophy of biology,
sociology, several collections of his metaphysical poems and plays, and even a
delightful little children's book!

\parindent -0.50cm

{\bf References:}

\baselineskip 0.45cm
\parskip 0.05cm

Christian, J. (1998) `Why the Quantum Must Yield to Gravity',
http://xxx.lanl.gov/abs/gr-qc/9810078. To appear in C. Callender and N.
Huggett (eds), {\it Physics Meets Philosophy at the Planck Scale}
(Cambridge: Cambridge University Press).

Penrose, R. (1996) `On Gravity's Role in Quantum State Reduction',
{\it General Relativity and Gravitation} {\bf 28}, 581-600.

Shimony, A. (1978)
`Metaphysical Problems in the Foundations of Quantum Mechanics',
{\it International Philosophical Quarterly} {\bf 18}, 3-17.

Shimony, A. (1998)
`Philosophical and Experimental Perspectives on Quantum Physics',
the 6$^{th}$ Vienna
Circle Lecture, Institute of Vienna Circle, Vienna. Unpublished.
\end